\documentclass[preprint]{JHEP3} 

\JHEPspecialurl{http://jhep.sissa.it/JOURNAL/JHEP3.tar.gz}
\usepackage{epsfig,multicol,bbm}
\usepackage{axodraw}

\newcommand\fverb{\setbox\fverbbox=\hbox\bgroup\verb}
\newcommand\fverbdo{\egroup\medskip\noindent%
            \fbox{\unhbox\fverbbox}\ }
\newcommand\fverbit{\egroup\item[\fbox{\unhbox\fverbbox}]}
\newbox\fverbbox

\title{Feynman Rules for the Rational Part of the
QCD 1-loop amplitudes}

\author{P. Draggiotis\\
       Departamento de F\'{i}sica Te\'orica y del Cosmos y CAFPE
       Universidad de Granada, E-18071 Granada, Spain.\\
       E-mail: \email{pdrangiotis@ugr.es}}

\author{M.V. Garzelli\\
       Departamento de F\'{i}sica Te\'orica y del Cosmos y CAFPE
       Universidad de Granada, E-18071 Granada, Spain and
       INFN Milano, I-20133 Milano, Italy. \\
       E-mail: \email{garzelli@to.infn.it}}

\author{C.G. Papadopoulos\\
       Institute of Nuclear Physics, NCSR Demokritos,
       15310 Athens, Greece. \\
       E-mail: \email{costas.papadopoulos@cern.ch}}

\author{R. Pittau\\
       Departamento de F\'{i}sica Te\'orica y del Cosmos y CAFPE
       Universidad de Granada, E-18071 Granada, Spain.\\
       E-mail: \email{pittau@ugr.es}}

\abstract{
 We compute the complete set of
Feynman Rules producing the Rational Terms of kind ${\rm R_2}$
needed to perform any QCD 1-loop calculation. We also explicitly
check that
in order to account for the entire ${\rm R_2}$ contribution, even in
case of processes with more than four external legs, only up to
four-point vertices are needed. Our results are expressed both in
the 't Hooft Veltman regularization scheme and in the Four
Dimensional Helicity scheme, using explicit color configurations as
well as the color connection language.}


\begin{document}

\newcounter{im}
\setcounter{im}{0}
\newcommand{\exampleSp}{\stepcounter{im}\includegraphics[scale=0.9]{SpinorExamples_\arabic{im}.eps}}
\newcommand{\myindex}[1]{\label{com:#1}\index{{\tt #1} & pageref{com:#1}}}
\renewcommand{\topfraction}{1.0}
\renewcommand{\bottomfraction}{1.0}
\renewcommand{\textfraction}{0.0}
\newcommand{\nn}{\nonumber \\}
\newcommand{\eqn}[1]{eq.~\ref{eq:#1}}
\newcommand{\be}{\begin{equation}}
\newcommand{\ee}{\end{equation}}
\newcommand{\ba}{\begin{array}}
\newcommand{\ea}{\end{array}}
\newcommand{\bea}{\begin{eqnarray}}
\newcommand{\eea}{\end{eqnarray}}
\newcommand{\bqa}{\begin{eqnarray}}
\newcommand{\eqa}{\end{eqnarray}}
\newcommand{\nl}{\nonumber \\}
\def\db#1{\bar D_{#1}}
\def\zb#1{\bar Z_{#1}}
\def\d#1{D_{#1}}
\def\tld#1{\tilde {#1}}
\def\slh#1{\rlap / {#1}}
\def\eqn#1{eq.~(\ref{#1})}
\def\eqns#1#2{Eqs.~(\ref{#1}) and~(\ref{#2})}
\def\eqnss#1#2{Eqs.~(\ref{#1})-(\ref{#2})}
\def\fig#1{Fig.~{\ref{#1}}}
\def\figs#1#2{Figs.~\ref{#1} and~\ref{#2}}
\def\sec#1{Section~{\ref{#1}}}
\def\app#1{Appendix~\ref{#1}}
\def\tab#1{Table~\ref{#1}}
\def\cg{c_\Gamma}
\newcommand{\bfig}{\begin{center}\begin{picture}}
\newcommand{\efig}[1]{\end{picture}\\{\small #1}\end{center}}
\newcommand{\flin}[2]{\ArrowLine(#1)(#2)}
\newcommand{\ghlin}[2]{\DashArrowLine(#1)(#2){5}}
\newcommand{\wlin}[2]{\DashLine(#1)(#2){2.5}}
\newcommand{\zlin}[2]{\DashLine(#1)(#2){5}}
\newcommand{\glin}[3]{\Photon(#1)(#2){2}{#3}}
\newcommand{\gluon}[3]{\Gluon(#1)(#2){5}{#3}}
\newcommand{\lin}[2]{\Line(#1)(#2)}
\newcommand{\sof}{\SetOffset}



\section{Introduction}
In the last few years, the problem of computing one-loop amplitudes
efficiently has been attacked by several groups. Standard
techniques, such as the Passarino-Veltman~\cite{Passarino:1978jh}
tensor reduction and its many variances~\cite{Denner:1991kt}-\cite{golem},
have been used for many years, and have produced a great deal
of useful results~\cite{classic_results}. Nowadays new developments,
in which the amplitude is directly reconstructed are widely used.
Such approaches rely on the fact that the basis of one-loop scalar
integrals is known in terms of Boxes, Triangles, Bubbles and (in
massive theories) Tadpoles, so that any one-loop amplitude ${\cal
M}$ can then be written as:
\bqa \label{master} {\cal M}= \sum_i d_i {\rm ~Box}_i +\sum_i c_i
{\rm ~Triangle}_i +\sum_i b_i {\rm ~Bubble}_i +\sum_i a_i {\rm
~Tadpole}_i + {\rm R}\,, \eqa
where $d_i$, $c_i$, $b_i$ and $a_i$ are the coefficients to be
determined and ${\rm R}$ is a remaining piece, called Rational Term
(RT). The first attempt in this direction was the Unitarity approach
~\cite{unitarity-cut}, by Bern, Dixon and Kososwer, in which
two-particle cuts are performed on the one-loop amplitude (or,
equivalently, two tree-level amplitudes are glued) in order to get
information on the coefficients in eq.~\ref{master}. The method
produced many useful -- mainly analytical -- results, especially for
massless theories~\cite{unitarity-results}, but a systematic way to
determine all the coefficients of eq.~\ref{master} was missing.

 Later it was shown by Britto Cachazo and Feng~\cite{Britto:2004nc}
that the $d_i$ coefficients
can be easily separated from the rest and computed by introducing quadruple
cuts
in which the loop integration momentum is completely frozen by four
on-shell conditions.
It was then possible to perform a full reconstruction of the amplitude
for theories with only boxes, such as N = 4  super-Yang-Mills.
However, a systematic procedure to get all the other coefficients
was still missing.

Recently the problem of determining in a systematic way the
coefficients $d_i$, $c_i$, $b_i$ and $a_i$ was completely solved by
the OPP method of refs.~\cite{opp} and~\cite{Ossola:2007ax}. Within
this method, eq.~\ref{master} is substituted by its unintegrated
counterpart, at the price of introducing the so called spurious
terms, defined by the property of vanishing upon integration over
the loop momentum $q$. In practice, since the functional form in $q$
of the spurious terms is universal, one has to find, besides $d_i$,
$c_i$, $b_i$, and $a_i$, an additional set of  coefficients.  The
OPP method allows to find all those coefficients by computing the
unintegrated amplitude at different values of $q$ for which 4, 3, 2
and 1 propagators vanish. At each stage, the coefficient that have
been already computed are numerically subtracted from the original
amplitude, so that, by using such an OPP subtraction, it is possible
to disentangle all the coefficients in a systematic way. The OPP
approach was inspired by the unitarity method and the tensor
reduction at the integrand level~\cite{intlevel}.

More recently the OPP subtraction method was used by the authors
of~\cite{Ellis:2007br} together with the  Unitarity approach, giving
rise to the so called generalized Unitarity techniques, that,
nowadays, include both semi-analytical~\cite{Berger:2008sj} and
fully numerical
versions~\cite{Giele:2008ve}-\cite{Lazopoulos:2008ex}. Nevertheless,
in practice, only the so called cut-constructible part of the
amplitude, namely that one proportional to the one-loop scalar
functions, can be easily obtained. The remaining Rational Terms
${\rm R}$~\cite{directcomp1}-\cite{directcomp2} require some
additional work. For instance, in~\cite{Giele:2008ve} the Rational
part is obtained by explicitly computing the amplitude at different
integer values of the space-time dimensions ($n$). Other
possibilities are to get them through $n$-dimensional
cuts~\cite{dcut} or with the help of recursion relations~\cite{rec}.

On the other hand, in the OPP method, two classes of terms
contributing to ${\rm R}$ naturally arise~\cite{Ossola:2008xq}. The
first class, called ${\rm R_1}$, can be derived straightforwardly
within the same framework used to determine all other coefficients,
while the second class, called ${\rm R_2}$, is  coming from the
$(n-4)$-dimensional part of the amplitude and can be obtained by
computing, once for all, tree-level like Feynman Rules for the
theory under study. Moreover, it is worthwhile to mention that only
the full ${\rm R= R_1+R_2}$ constitutes a physical gauge-invariant
quantity in dimensional regularization. On the other hand, ${\rm
R_1}$ can be directly read out from the analytic expressions of the
cut-constructible part of the amplitude, irrespectively of the
method used to derive it.

  In this paper, we explicitly compute the entire set of Feynman Rules
producing ${\rm R_2}$ needed in any (massive or massless)
QCD 1-loop calculation. We perform our calculation in the 
$\xi = 1$ 't Hooft-Feynman gauge. As a consequence, also the 
$\delta Z$ couterterms~\cite{Denner:1991kt} needed to build 
renormalized scattering amplitudes should be computed in the $\xi = 1$ gauge.

In the next section we briefly recall the origin of
${\rm R_2}$ and give a detailed computational example. In section~\ref{results},
we list our results and present numerical comparisons with known amplitudes.
In section~\ref{conclusions} we draw our conclusions and, in three appendices,
we collect diagrams and formulae used for the calculation as well
as our results expressed in the color connection language.

\section{The origin of ${\rm R_2}$ \label{origin}}
Before carrying out our program, we spend a few more words on
the origin of ${\rm R_2}$, that is also necessary for setting up the framework
of our calculation.

Our starting point is the general expression for the
{\it integrand} of a generic $m$-point
one-loop (sub-)amplitude
\bqa
\label{eq:1}
\bar A(\bar q)= \frac{\bar N(\bar q)}{\db{0}\db{1}\cdots \db{m-1}}\,,~~~
\db{i} = ({\bar q} + p_i)^2-m_i^2\,,
\eqa
where ${\bar q}$ is the integration momentum.
In the previous equation, dimensional regularization is assumed, so that
we use a bar to denote objects living
in $n=~4+\epsilon$ dimensions and a tilde to represent $\epsilon$-dimensional
quantities.
Notice that, when a $n$-dimensional index
is contracted with a 4-dimensional (observable) vector
$v_\mu$, the $4$-dimensional part is automatically selected. For example
\bqa
\label{noeps}
\bar q \cdot v \equiv (q+ {\tld q}) \cdot v\,= q \cdot v\,~~~{\rm and}~~~
\rlap/ {\bar v} \equiv  \bar \gamma_{\bar \mu}\, v^\mu = \rlap /v\,.
\eqa
An important consequence is
\bqa
{\bar q}^2= q^2 + {\tld q}^2\,.
\eqa
The numerator function $\bar{N}(\bar q)$ can be further
split into a $4$-dimensional plus an $\epsilon$-dimensional part
\bqa
\label{eq:split}
\bar{N}(\bar q) = N(q) + \tld{N}(\tld{q}^2,q,\epsilon)\,.
\eqa
$N(q)$ lives in $4$-dimensions while
$\tld{N}(\tld{q}^2,q,\epsilon)$, once integrated,
gives rise to the RTs of kind ${\rm R_2}$,
defined as
\bqa
\label{eqr2}
{\rm R_2} \equiv  \frac{1}{(2 \pi)^4}\int d^n\,\bar q
\frac{\tld{N}(\tld{q}^2,q,\epsilon)}{\db{0}\db{1}\cdots \db{m-1}} \,.
\eqa
To investigate the explicit form of $\tld{N}(\tld{q}^2,q,\epsilon)$
it is important to understand better the separation in eq.~\ref{eq:split}.
From a given {{\em integrand} $\bar A (\bar q)$
this is obtained by splitting, in the numerator function, the $n$-dimensional integration momentum
${\bar q}$, the $n$-dimensional gamma matrices
$\bar \gamma_{\bar \mu}$  and the $n$-dimensional metric tensor
$\bar g^{\bar \mu \bar \nu}$ into a $4$-dimensional
component plus remaining pieces:
\bqa
\label{qandg}
\bar q                 &=& q + \tld{q}\,, \nl
\bar \gamma_{\bar \mu} &=&  \gamma_{\mu}+ \tld{\gamma}_{\tld{\mu}}\,,\nl
 \bar g^{\bar \mu \bar \nu}  &=&  g^{\mu \nu}+  \tld{g}^{\tld{\mu} \tld{\nu}}\,.
\eqa

A practical way to determine ${\rm R_2}$ is then computing, once for all
and with the help of eq.~\ref{qandg}, tree-level like Feynman Rules
by calculating the ${\rm R_2}$ part coming from
one-particle irreducible amplitudes up to four external legs.
The fact that four external legs are enough is guaranteed
by the ultraviolet nature of the RTs, proven in~\cite{directcomp1}.
Through eq.~\ref{eqr2} a set of basic integrals with up to 4 denominators is generated,
containing powers of $\tld{q}$ and $\epsilon$ in the numerator. A list that exhausts all possibilities in the $\xi = 1$ 't Hooft-Feynman gauge 
is presented in appendix~\ref{appa}.
Notice that, according to the chosen regularization scheme,
results may differ. In eq.~\ref{eqr2} we use the
't Hooft-Veltman (HV) scheme, while in the Four Dimensional Helicity scheme
(FDH), any explicit $\epsilon$ dependence in the numerator function
is discarded before integration. Therefore
\bqa
\label{eqr2fdh}
{\rm R_2} \Bigl |_{FDH} =  \frac{1}{(2 \pi)^4}\int d^n\,\bar q
\frac{\tld{N}(\tld{q}^2,q,\epsilon= 0)}{\db{0}\db{1}\cdots \db{m-1}} \,.
\eqa

As an explicit and simple example of the described procedure,
we detail the calculation of
${\rm R_2}$ coming from the gluon self-energy.
The contributing diagrams \footnote{Our conventions and notations are listed in~\app{appb}.}
are drawn in fig.~\ref{selfg}.

As for the ghost loop with 2 external gluons, we can write
the numerator as
\begin{eqnarray}
\bar{N}(\bar{q}) = \frac{g^2}{(2\pi)^4} f^{a_1bc}\,
f^{a_2cb}\,(p+\bar{q})^{\mu_1}
\bar{q}^{\mu_2}
\,.
\end{eqnarray}
Since $\mu_1$ and $\mu_2$ are external Lorentz indices, that are
eventually contracted with 4-dimensional external currents, their
$\epsilon$-dimensional component is killed due to
eq.~\ref{noeps}. Therefore, ${\rm R_2}= 0$ for this diagram, being
$\tld{N}(\tld{q}^2,q,\epsilon)= 0$.
With this same reasoning, one easily shows that
ghost loops never contribute to ${\rm R_2}$, even with 3 or 4
external gluons.

The contribution due to $N_f$ quark loops is given by
the second diagram of fig.~\ref{selfg}, whose numerator reads
\bqa
\bar N(\bar q) &=& -\frac{g^2}{(2 \pi)^4} N_f \frac{\delta_{a_1a_2}}{2}\,
Tr[{\gamma}^{{\mu_1}}(\rlap/{\bar{q}}+m_q) {\gamma}^{{\mu_2}}
(\rlap/{\bar{q}}+\rlap/{p}+m_q)]\,,
\eqa
where the external indices $\mu_1$ and $\mu_2$ have been directly taken
in 4 dimensions.
By anti-commuting ${\gamma}^{{\mu_2}}$ and $\rlap/{\bar{q}}$ and using the fact that,
due to Lorentz invariance, odd powers of ${\tld{q}}$ do no contribute, one immediately arrives
at the result
\bqa
\label{ntilda2}
\tld{N}({\tld{q}}^2)= \frac{g^2}{8 \pi^4}\, N_f\,
\delta_{a_1a_2}\,
g_{\mu_1 \mu_2} {\tld q}^2\,.
\eqa
Eq.~\ref{ntilda2}, integrated with the help of the first one of
eqs.~\ref{2int}, gives
the term proportional to $N_f$ in the 2-point effective vertex of
fig.~\ref{effectivevertices}.

\begin{figure}[t]
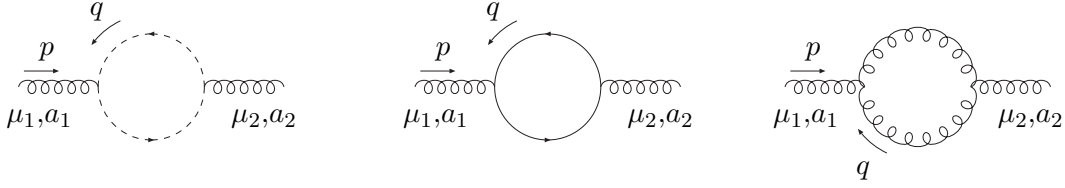

\bfig(300,70)
\SetScale{0.5}
\sof(-70,0)
\DashArrowArc(150,50)(40,0,180){5}
\DashArrowArc(150,50)(40,180,360){5}
\Gluon(50,50)(110,50){5}{5}
\Gluon(190,50)(250,50){5}{5}
\LongArrow(54,62)(79,62)
\LongArrowArc(150,50)(55,115,145)
\Text(33,35)[bl]{$p$}
\Text(52,50)[bl]{$q$}
\Text(33,13)[]{$\mu_1$,$a_1$}
\Text(118,13)[]{$\mu_2$,$a_2$}
\sof(80,0)
\ArrowArc(150,50)(40,0,180)
\ArrowArc(150,50)(40,180,360)
\Gluon(50,50)(110,50){5}{5}
\Gluon(190,50)(250,50){5}{5}
\LongArrow(54,62)(79,62)
\LongArrowArc(150,50)(55,115,145)
\Text(33,35)[bl]{$p$}
\Text(52,50)[bl]{$q$}
\Text(33,13)[]{$\mu_1$,$a_1$}
\Text(118,13)[]{$\mu_2$,$a_2$}
\sof(220,0)
\GlueArc(150,50)(40,0,180){5}{8}
\GlueArc(150,50)(40,180,360){5}{8}
\Gluon(50,50)(110,50){5}{5}
\Gluon(190,50)(250,50){5}{5}
\LongArrow(54,62)(79,62)
\LongArrowArcn(150,50)(55,-115,-145)
\Text(33,35)[bl]{$p$}
\Text(52,-10)[bl]{$q$}
\Text(33,13)[]{$\mu_1$,$a_1$}
\Text(118,13)[]{$\mu_2$,$a_2$}
\end{picture}
\end{center}
\caption{\em Diagrams contributing to the gluon self-energy.}
\label{selfg}
\end{figure}
\begin{center}
\begin{figure}[ht]
\bfig(300,500)
\SetScale{0.5}
\sof(-50,440)
\LongArrow(47,62)(73,62)
\Text(27,35)[bl]{$p$}
\Text(17,15)[]{$\mu_1$,$a_1$}
\Text(68,15)[]{$\mu_2$,$a_2$}
\gluon{35,50}{85,50}{4}
\gluon{85,50}{135,50}{4}
\GCirc(85,50){6}{0}
\Text(90,25)[l]{$\displaystyle = \frac{i g^2 N_{col}}{48 \pi^2} \,
\delta_{a_1a_2}\, \Bigl[\,\frac{p^2}{2} g_{\mu_1\mu_2}
+\lambda_{HV}\,\Bigl( g_{\mu_1\mu_2} p^2-p_{\mu_1} p_{\mu_2}\Bigr) $}
\Text(170,-10)[l]{$\displaystyle
+\frac{N_f}{N_{col}}\,(p^2-6\, m_q^2)\,g_{\mu_1\mu_2}\Bigr]$}
\sof(-50,340)
\gluon{130,5}{85,50}{5}\gluon{85,50}{130,95}{5}
\gluon{20,50}{85,50}{5}
\Text(46,49)[]{$p_2$}
\Text(25,40)[]{$p_1$}
\Text(45,3)[]{$p_3$}
\Text(75,55)[]{$\mu_2$,$a_2$}
\Text(7,15)[]{$\mu_1$,$a_1$}
\Text(75,-6)[]{$\mu_3$,$a_3$}
\LongArrow(35,62)(66,62)
\LongArrow(105,10)(85,30)
\LongArrow(110,95)(90,75)
\GCirc(85,50){6}{0}
\Text(90,25)[l]{$\displaystyle = -\frac{g^3 N_{col}}{48 \pi^2}\,
\left(\frac{7}{4}+\lambda_{HV} +2 \frac{N_f}{N_{col}} \right)
 \,
f^{a_1 a_2 a_3}\, V_{\mu_1\mu_2\mu_3}(p_1,p_2,p_3)$}
%
%
\sof(-50,240)
\gluon{130,5}{85,50}{5}\gluon{85,50}{130,95}{5}
\gluon{40,5}{85,50}{5}\gluon{85,50}{40,95}{5}
\Text(68,0)[l]{\small $\mu_3$,$a_3$}
\Text(17,0)[r]{\small $\mu_4$,$a_4$}
\Text(68,50)[l]{\small $\mu_2$,$a_2$}
\Text(17,50)[r]{\small $\mu_1$,$a_1$}
\GCirc(85,50){6}{0}
\Text(90,25)[l]{$\displaystyle = -\frac{i g^4 N_{col}}{96\pi^2}\,
\sum_{P(234)}
\left\{ \Bigl[\,\frac{
    \delta_{a_1 a_2}\delta_{a_3 a_4}+
    \delta_{a_1 a_3}\delta_{a_4 a_2}+
    \delta_{a_1 a_4}\delta_{a_2 a_3}}{N_{col}}\right.$ }
\Text(160,-10)[l]{$\displaystyle
      +\, 4\, Tr(t^{a_1}t^{a_3}t^{a_2}t^{a_4}+
                    t^{a_1}t^{a_4}t^{a_2}t^{a_3})\,(3+\lambda_{HV})$}
\Text(160,-45)[l]{$\displaystyle
      -\,  Tr(\{t^{a_1}t^{a_2}\}\{t^{a_3}t^{a_4}\})
        \,(5+2 \lambda_{HV})
    \Bigr ]\,g_{\mu_1\mu_2} g_{\mu_3\mu_4}$}
\Text(90,-80)[l]{$\displaystyle \left. + 12 \frac{N_f}{N_{col}}
Tr(t^{a_1}t^{a_2}t^{a_3}t^{a_4})
\left(\frac{5}{3} g_{\mu_1\mu_3} g_{\mu_2\mu_4}
           - g_{\mu_1\mu_2} g_{\mu_3\mu_4}
           - g_{\mu_2\mu_3} g_{\mu_1\mu_4}\right)
\right\}$}
\sof(-50,10)
\flin{130,5}{85,50}\flin{85,50}{130,95}
\gluon{20,50}{85,50}{5}
\Text(7,20)[r]{\small $\mu,a $}
\Text(71,51)[]{$k$}
\Text(71,0)[]{$l$}
\GCirc(85,50){6}{0}
\Text(90,25)[l]{$\displaystyle = \frac{i g^3}{16 \pi^2 }
\frac{N_{col}^2-1}{2 N_{col}}\,t_{kl}^a\gamma_\mu\,(1+ \lambda_{HV})$}
\sof(-50,70)
\LongArrow(47,59)(73,59)
\Text(27,32)[bl]{$p$}
\flin{35,50}{85,50}
\GCirc(85,50){6}{0}
\flin{85,50}{135,50}
\Text(17,15)[]{$l$}
\Text(68,15)[]{$k$}
\Text(90,25)[l]{$\displaystyle = \frac{i g^2}{16 \pi^2}
\frac{N_{col}^2-1}{2 N_{col}}\delta_{kl}
(-\rlap/p + 2 m_q)\,\lambda_{HV}$}
\end{picture}
\end{center}
\caption{\em Effective vertices contributing to ${\rm R_2}$ in pure QCD.
$\sum_{P(234)}$ stands for a summation over the six
permutations of the indices 2, 3 and 4, and $\{t^{a_i}t^{a_j}\}
\equiv t^{a_i} t^{a_j}+t^{a_j} t^{a_i}$.
$\lambda_{HV}= 1$ in the HV  scheme and
$\lambda_{HV}= 0$ in the FDH scheme.
$N_{col}$ is the number of colors and $N_f$ is the number of fermions running
in the quark loop.}
\label{effectivevertices}
\end{figure}
\end{center}
\clearpage
\begin{center}
\begin{figure}[ht]
\bfig(300,560)
\SetScale{0.5}
\sof(-50,484)
\flin{130,5}{85,50}\flin{85,50}{130,95}
\glin{20,50}{85,50}{8}
\Text(7,20)[r]{\small $\mu$}
\Text(22,29)[b]{\small $V$}
\Text(71,51)[]{$k$}
\Text(71,0)[]{$l$}
\GCirc(85,50){6}{0}
\Text(90,25)[l]{$\displaystyle = -\frac{g^2}{16 \pi^2 }
\frac{N_{col}^2-1}{2 N_{col}}\,\delta_{kl} \gamma_\mu\,
(v + a \gamma_5)\,
(1+ \lambda_{HV})$}
\sof(-50,408)
\flin{130,5}{85,50}\flin{85,50}{130,95}
\DashLine(20,50)(85,50){2}
\Text(22,29)[b]{\small $S$}
\Text(71,51)[]{$k$}
\Text(71,0)[]{$l$}
\GCirc(85,50){6}{0}
\Text(90,25)[l]{$\displaystyle = -\frac{g^2}{8 \pi^2 }
\frac{N_{col}^2-1}{2 N_{col}}\,\delta_{kl} \,
(c + d \gamma_5)\,
(1+ \lambda_{HV})$}
\sof(-50,331)
\gluon{130,5}{85,50}{5}\gluon{85,50}{130,95}{5}
\glin{20,50}{85,50}{8}
\Text(7,20)[r]{\small $\mu$}
\Text(22,29)[b]{\small $V$}
\Text(46,49)[]{$p_1$}
\Text(45,3)[]{$p_2$}
\LongArrow(105,10)(85,30)
\LongArrow(110,95)(90,75)
\Text(79,51)[]{\small $\alpha_1$,$a_1$}
\Text(79,-3)[]{\small $\alpha_2$,$a_2$}
\GCirc(85,50){6}{0}
\Text(90,25)[l]{$\displaystyle = -a\,\frac{ig^2}{12 \pi^2 } \,\delta_{a_1 a_2}
\epsilon_{\mu\alpha_1\alpha_2\beta}\, (p_1-p_2)^{\beta}$}
\sof(-50,256)
\gluon{130,5}{85,50}{5}\gluon{85,50}{130,95}{5}
\DashLine(20,50)(85,50){2}
\Text(79,51)[]{\small $\alpha_1$,$a_1$}
\Text(79,-3)[]{\small $\alpha_2$,$a_2$}
\Text(22,29)[b]{\small $S$}
\GCirc(85,50){6}{0}
\Text(90,25)[l]{$\displaystyle = c\,\frac{g^2}{8 \pi^2 } \, \delta_{a_1 a_2}
g_{\alpha_1 \alpha_2}\,m_q $}
\sof(-50,180)
\gluon{130,5}{85,50}{5}\gluon{85,50}{130,95}{5}
\glin{40,5}{85,50}{8}\glin{85,50}{40,95}{8}
\Text(68,0)[l]{\small $\alpha_2$,$a_2$}
\Text(17,0)[r]{\small $\mu_2$}
\Text(68,50)[l]{\small $\alpha_1$,$a_1$}
\Text(17,48)[r]{\small $\mu_1$}
\Text(37,53)[r]{\small $V_1$}
\Text(37,-3)[r]{\small $V_2$}
\GCirc(85,50){6}{0}
\Text(90,25)[l]{$\displaystyle = -\frac{i g^2}{24 \pi^2}
\,\delta_{a_1 a_2}
(v_1 v_2+a_1a_2)\,
( g_{\mu_1\mu_2} g_{\alpha_1\alpha_2}
 +g_{\mu_1\alpha_1} g_{\mu_2\alpha_2}
 +g_{\mu_1\alpha_2} g_{\mu_2\alpha_1})$}
\sof(-50,104)
\gluon{130,5}{85,50}{5}\gluon{85,50}{130,95}{5}
\DashLine(40,5)(85,50){2}
\DashLine(85,50)(40,95){2}
\Text(68,0)[l]{\small $\alpha_2$,$a_2$}
\Text(68,50)[l]{\small $\alpha_1$,$a_1$}
\Text(37,50)[r]{\small $S_1$}
\Text(37,0)[r]{\small $S_2$}
\GCirc(85,50){6}{0}
\Text(90,25)[l]{$\displaystyle = \frac{i g^2}{8 \pi^2}
\,\delta_{a_1 a_2}\,(c_1 c_2-d_1d_2)\,g_{\alpha_1\alpha_2}$}
\sof(-50,28)
\gluon{130,5}{85,50}{5}\gluon{85,50}{130,95}{5}
\gluon{40,5}{85,50}{5}\glin{85,50}{40,95}{8}
\Text(17,0)[r]{\small $\alpha_3$,$a_3$}
\Text(68,0)[l]{\small $\alpha_2$,$a_2$}
\Text(17,48)[r]{\small $\mu$}
\Text(68,50)[l]{\small $\alpha_1$,$a_1$}
\Text(37,50)[r]{\small $V$}
\GCirc(85,50){6}{0}
\Text(90,25)[l]{$\displaystyle =  -\frac{g^3}{24 \pi^2}
\left\{v \,Tr(t^{a_1}\{t^{a_2}t^{a_3}\})
( g_{\mu\alpha_1} g_{\alpha_2\alpha_3}
 +g_{\mu\alpha_2} g_{\alpha_1\alpha_3}
 +g_{\mu\alpha_3} g_{\alpha_1\alpha_2}) \right.$}
\Text(144,-10)[l]{$\displaystyle \left.
- i9a\, \left[Tr(t^{a_1}t^{a_2}t^{a_3})-Tr(t^{a_1}t^{a_3}t^{a_2})\right]\,
\epsilon_{\mu \alpha_1\alpha_2\alpha_3}
\right\}$}
\end{picture}
\end{center}
\caption{\em Effective vertices contributing to ${\rm R_2}$ in mixed QCD.
$\lambda_{HV}= 1$ in the HV   scheme and
$\lambda_{HV}= 0$ in the FDH  scheme.
In the case of neutral external vectors or scalars, the formulae should 
be read as the contribution given by one quark loop. 
In the case of charged external particles, they refer instead 
to the contribution of one quark family.}
\label{effectivemixvertices}
\end{figure}
\end{center}
Finally, the numerator function of the diagram with a gluonic loop reads
\begin{eqnarray}
\bar N(\bar q) = - \frac{g^2}{2 (2\pi)^4}  f^{a_1bc}\,
f^{a_2cb} V_{\mu_1\bar \beta\bar \gamma}(p, -\bar q -p, \bar q)
V_{\mu_2}^{~\,\bar \gamma \bar \beta}
(-p,-\bar q,\bar q + p)\,,
\end{eqnarray}
with $V$ given in eq.~\ref{eq:vdef}.
The contraction of the two $V$ tensors gives terms containing
\bqa
{\bar q}^2 = q^2 + {\tld q}^2\,,~~{\rm and}~~
\bar g_{\bar \alpha \bar \beta}\, \bar g^{\bar \alpha \bar \beta}= 4 + \epsilon\,,
\eqa
from which one obtains
\begin{eqnarray}
\label{ggg}
\tld{N}({\tld{q}}^2, q, \epsilon) =
- \frac{g^2} {2 (2\pi)^4}   f^{a_1bc}
f^{a_2cb} \Bigl[\,2 g_{\mu_1\mu_2} \tld{q}^2
+ 4\epsilon q_{\mu_1} q_{\mu_2} + 2 \epsilon q_{\mu_1} p_{\mu_2} + 2 \epsilon
p_{\mu_1} q_{\mu_2}
 + \epsilon p_{\mu_1} p_{\mu_2}\,\Bigr] \,. \nonumber \\
\end{eqnarray}
Performing the integration, gives the expression written in the first
line of fig.~\ref{effectivevertices}, where the contribution proportional to $\lambda_{HV}$
is generated by the $\epsilon$ dependence of eq.~\ref{ggg}.

The complete set of effective vertices obtained with the described technique is presented in the next section.
\section{Results\label{results}}
As already explained, 1-loop irreducible Feynman diagrams
up to 4 external legs are sufficient to compute ${\rm R_2}$ for any amplitude
with any number of external legs.
Each contributing diagram has been calculated analytically
by using the Feynman rules listed in appendix~\ref{appb}, which
also contains the list of the relevant graphs.
Different contributions have then been
summed and reorganized to identify the effective
${\rm R_2}$ vertices listed in
figs.~\ref{effectivevertices} and ~\ref{effectivemixvertices}, which
represent the main result of this work and that allow to determine
${\rm R_2}$ needed in the computation of the NLO QCD corrections
to any process in the Standard Model.

In fig.~\ref{effectivevertices} we collect the ``pure''
QCD effective vertices, namely all vertices generated by QCD corrections
to processes with external QCD particles. The complete set
of contributing diagrams  is given in fig.~\ref{loopdiagrams}.
In fig.~\ref{effectivemixvertices} we list, instead,
the ``mixed'' QCD vertices
generated by QCD corrections to processes containing at least one
external electroweak particle. In this paper, this last class is parametrized
by introducing generic couplings of a (pseudo)-vector
and of a (pseudo)-scalar with a quark line, as in the last
two vertices of fig.~\ref{qcdrules}.
The non vanishing diagrams contributing to  ${\rm R_2}$ are listed in
fig.~\ref{loopmixdiagrams}.

In all figures, $N_{col}$ is the number of colours,
$N_f$ is the number of fermions running in the quark loop and
$\lambda_{HV}$ is a parameter allowing to read our formulae
in two different regularization schemes:
$\lambda_{HV}= 1$ corresponds to the HV scheme
of eq.~\ref{eqr2} while $\lambda_{HV}= 0$ in the FDH
scheme defined in eq.~\ref{eqr2fdh}.

Notice that the whole structure of the three-gluon effective vertex
is always proportional to the tree-level, while the four-gluon effective
vertex is more complicated.

Notice also that, when a completely antisymmetric $\epsilon$ tensor occur in the formulae of
the mixed QCD vertices, it always multiplies the axial coupling $a$.
Therefore, summing over all quark loops gives zero in the Standard Model, due
to the anomaly cancellation. Such terms can then be taken to be zero
from the very beginning.

In figs.~\ref{effectivevertices} and~\ref{effectivemixvertices}
the effective vertices  are given
in terms of traces of color matrices and structure constants.
The same result in terms of color connections is presented in
appendix~\ref{appc}.

Just as a showcase of our ability to reproduce the rational terms ${\rm R_2}$ correctly for higher multiplicity
of external legs, we have computed the 1-loop six gluon amplitude using an extension \cite{vanHameren:2009dr}
of {\tt HELAC-PHEGAS} \cite{Kanaki:2000ey},   {\tt HELAC-1loop},   that includes virtual corrections through an interface with
{\tt CutTools} \cite{Ossola:2007ax}.

The comparison against the results of ref.~\cite{Giele:2008bc} 
is given in table~\ref{table1}, that refers to
one color configuration only and to the following phase space point:

\begin{footnotesize}
\begin{eqnarray}
\label{specif}
p_1&=&(-3.000000000000000, 1.837117307087384,-2.121320343559642, 1.060660171779821)\nn
p_2&=&(-3.000000000000000,-1.837117307087384, 2.121320343559642,-1.060660171779821)\nn
p_3&=&( 2.000000000000000, 0.000000000000000,-2.000000000000000, 0.000000000000000)\nn
p_4&=&( 0.857142857142857, 0.000000000000000, 0.315789473684211, 0.796850604480708)\nn
p_5&=&( 1.000000000000000, 0.866025403784439, 0.184210526315789, 0.464829519280413)\nn
p_6&=&( 2.142857142857143,-0.866025403784439, 1.500000000000000,-1.261680123761121)
\end{eqnarray}
\end{footnotesize}
We find an excellent agreement among the results when including the
${\rm R_2}$ contribution. Finally, we mention that, always with the
help of {\tt HELAC-1loop}, we successfully compared our predictions
for six-quark 1-loop amplitudes (with three different flavours) with the
results produced by the GOLEM group~\cite{golem}. 
In addition, for all sub-processes 
included in the 2007 Les Houches wish list~\cite{Bern:2008ef},
we explicitly checked that the validity of the Ward Identity 
for a single on-shell external gluon (when present) 
is preserved by the sum ${\rm R_1}+{\rm R_2}$.
\begin{table}
\begin{scriptsize}
\begin{tabular}{|c|c|c|c|}
\hline
CC & ${\rm R_1}$ & ${\rm R_2}$ & $|A_{6}|$ \\
\hline \hline
\multicolumn{4}{|c|}{$++++++$}
\\ \hline
\multicolumn{4}{|c|}{$2.417991007837614\cdot 10^ {-17}
-i~4.793031572579084\cdot 10^ {-17}$}
\\ \hline
$-$ & $-$ & $-$ &  $ 0.529806483643855$
\\
$-$ & $-$&$-$ & $0.529806483661295$
\\
$4.105695207224680\cdot 10^{-16}$ & $ -0.1382337836650667 $
&$-9.500309796373058\cdot 10^ {-2} $
&$0.529806483661327$
\\
$+i~1.504282270415678\cdot 10^{-15}$ & $+i~0.360080242802749$
& $+i~0.1156249885933891$  & \\
\hline \hline
\multicolumn{4}{|c|}{$-+++++$}
\\ \hline
\multicolumn{4}{|c|}{$9.755813312628327,\cdot 10^ {-19}
+i~1.353208509225273\cdot 10^ {-16}$}
\\ \hline $-$
& $-$ &$-$& $3.25996704351899$
\\ $-$
&  $-$ &$-$&  $3.25996705427236$
\\
$3.812889817765093\cdot 10^ {-10}$
& $1.762222300120013$
& $-3.777721263438941\cdot
10^{-2}$
& $3.25996705427262$
\\
$+i~ 1.011089409856760\cdot 10^ {-9}$
&$+i~ 2.633781480837803$
&$+i~ 0.1327460047079652$
&
\\
\hline \hline
\multicolumn{4}{|c|}{$--++++$}
\\ \hline
\multicolumn{4}{|c|}{$-2.806856204696471 -i~28.35268397049988$}
\\ \hline $-$
& $-$&$-$&$   1373.74753500854$
\\ $-$
& $-$&$-$ &$1373.74753500828$
\\
$33436785.84436276$
& $-33438045.33273086$
&$ 6.674685589922374$
&$1373.74753500852$
\\
$-i~14771512.10073091$
& $+i~14772021.20888190$
&$+i~61.85407887654488$   &
\\
\hline \hline
\multicolumn{4}{|c|}{$-+-+-+$}
\\ \hline
\multicolumn{4}{|c|}{$3.131859164936308- i~0.2073463331363808$}
\\ \hline $-$
&  $-$& $-$ &  $ 151.043950328960$
\\ $-$
& $-$ & $-$ & $151.043950337947$
\\
$-370592.0271294174$
& $ 370518.3987124019$
& $ -6.563948999154568   $
& $151.043950337955$
\\
$+i~28580469.70894824$
&$-i~28580599.25896605 $
&$+i~ 0.9730930522347396 $
&
\\
\hline \hline
\multicolumn{4}{|c|}{$+-+-+-$}
\\ \hline
\multicolumn{4}{|c|}{$3.131859164936308+ i~0.2073463331363808$}
\\ \hline $-$
& $-$&$-$ & $ 153.780101529836$
\\ $-$
& $-$ & $-$ & $ 153.780101415986 $
\\
$-370533.5799977572$
&$ 370518.3987124019$
&$ -6.563948999154568   $
& $ 153.780101615242 $
\\
$-i~28580750.73535662$
&  $+i~28580599.25896605 $
& $-i~ 0.9730930522347396 $
&
\\
\hline
\end{tabular}
\caption{ \label{table1}
Results for the finite part of the 1-loop
virtual amplitudes for some helicity configurations for the case of
six external gluons for the phase space point given in the text. The
first row for each helicity configuration is the tree-order result.
The second (unitarity) and the third (semi-numerical) rows are the
results for $|A_{6}|$ taken from \cite{Giele:2008bc}. The fourth is
our result for the cut-constructible, CC (with renormalization scale
$\mu=\sqrt{s}$), $R_1$ and $R_2$ terms in the HV scheme as well as
for the $|A_{6}|$ in FDH scheme, to facilitate comparisons. The
relation of HV scheme result to the FDH scheme result is given in
\cite{Giele:2008bc}. }
\end{scriptsize}
\end{table}

\section{Conclusions\label{conclusions}}
We have derived the tree-level Feynman rules
needed to compute the Rational Terms ${\rm R_2}$ in QCD, both
using explicit color configurations and in the color connection language.
We listed all effective vertices generated by QCD corrections
to processes with external QCD particles and all possible mixed QCD effective vertices generated by QCD corrections to processes with at least one external EW
particle.
The inclusion of the derived vertices in an actual calculation gives
numerical agreement with known expressions
for processes up to 6 external legs. So we have explicitly checked  that
1,2,3 and 4-point vertices are enough to solve the problem for
an arbitrary number of external legs.
 In addition, all relevant integrals needed to compute ${\rm R_2}$
in the $\xi = 1$ 't Hooft-Feynman gauge have been explicitly listed.
The next obvious step is the determination of the Feynman Rules
needed in the complete Standard Model.
We leave this for a future publication.

\section*{Acknowledgments}

We would like to thank Giovanni Ossola and Thomas Binoth for many
useful discussions.
R.P. acknowledges the financial support of the ToK Program
``ALGOTOOLS'' (MTKD-CD-2004-014319). P.D.'s, C.G.P.'s and R.P.'s
research was partially supported by the RTN European Programme
MRTN-CT-2006-035505 (HEPTOOLS, Tools and Precision Calculations for
Physics Discoveries at Colliders). The research of R.P., M.V.G. and
P.D. was also supported by the MEC project FPA2008-02984.

\section*{Appendices}
\appendix
\section{The needed integrals \label{appa}}
In this appendix, we collect the integrals
needed to perform our calculation of ${\rm R_2}$. 
A non vanishing contribution is generated
only by integrals with zero or higher superficial degree of divergence.
They fall in two classes, namely
integrals involving even powers of $\tld{q}$ (odd powers do no contribute
due to Lorentz invariance) and Pole Parts (P.P.) of ultraviolet divergent integrals.
This second class is relevant when using regularization schemes
(such as the HV one) where the $\epsilon$
dependence in the numerator is kept.
In the following, we further classify the integrals according to the
number of denominators $\db{i} = ({\bar q} + p_i)^2-m_i^2$.
 The results for the Pole Parts have been checked against 
ref.~\cite{Denner:1991kt}.

\subsubsection*{2-point integrals:}
\begin{eqnarray}
\label{2int}
\int d^n\bar{q} \frac{\tld{q}^2}{\db{i}\db{j}}  & = & - \frac{i\pi^2}{2}
\left[m_i^2 + m_j^2 - \frac{(p_i-p_j)^2}{3}\right] 
+ O (\epsilon) \,, \nonumber \\
P.P. \left(\int d^n\bar{q} \frac{1}{\db{i}\db{j}} \right)
& = & -2 \frac{i\pi^2}{\epsilon} \, , \nonumber \\
P.P. \left(\int d^n\bar{q} \frac{q_\mu}{\db{i}\db{j}} \right)
& = & \frac{i\pi^2}{\epsilon}(p_i+p_j)_\mu \, ,\nonumber \\
P.P. \left(\int d^n\bar{q} \frac{q_\mu q_\nu}{\db{i}\db{j}} \right)
& = &
\frac{i\pi^2}{3\epsilon} \Bigl\{\,
\frac{(p_i-p_j)^2 -3 m_i^2 - 3 m_j^2 }{2} g_{\mu\nu}
-2\,p_{i\mu} p_{i\nu}
-2\,p_{j\mu} p_{j\nu}  \nonumber \\
&~&  
 - p_{i\mu} p_{j\nu}
 - p_{j\mu} p_{i\nu} \Bigr\}\,.
\end{eqnarray}
\subsubsection*{3-point integrals:}
\begin{eqnarray}
\label{3int}
\int d^n\bar{q} \frac{\tld{q}^2}{\db{i}\db{j}\db{k}}
& = & - \frac{i\pi^2}{2} + O (\epsilon) \, , \nonumber \\
\int d^n\bar{q} \frac{\tld{q}^2 q_\mu}{\db{i}\db{j}\db{k}}  & = &
\frac{i\pi^2}{6}(p_{ijk})_\mu  +
O (\epsilon) \, , \nonumber \\
P.P. \left(\int d^n\bar{q}
\frac{q_\mu q_\nu }{\db{i}\db{j}\db{k}}\right)  & = &
- \frac{i\pi^2}{2 \epsilon} g_{\mu\nu} \, , \nonumber \\
P.P. \left(\int d^n\bar{q} \frac{q_\mu q_\nu q_\rho}{\db{i}\db{j}\db{k}}
\right)
& = & \frac{i\pi^2}{6 \epsilon}\, {\Bigl [} g_{\mu\nu}(p_{ijk})_\rho +
g_{\nu\rho}(p_{ijk})_\mu + g_{\mu\rho}(p_{ijk})_\nu
\Bigr ] \,,
\end{eqnarray}
with $p_{ijk}= p_i+p_j+p_k$.
\subsubsection*{4-point integrals:}
\begin{eqnarray}
\label{4int}
\int d^n\bar{q} \frac{\tld{q}^4}{\db{i}\db{j}\db{k}\db{l}}  & = &
- \frac{i\pi^2}{6}  + O (\epsilon) \, , \nonumber \\
\int d^n\bar{q} \frac{\tld{q}^2q_{\mu}q_{\nu}}{\db{i}\db{j}\db{k}\db{l}}  & = &
- \frac{i\pi^2}{12} g_{\mu\nu} + O (\epsilon) \, ,\nonumber \\
\int d^n\bar{q} \frac{\tld{q}^2q^2}{\db{i}\db{j}\db{k}\db{l}}  & = & - \frac{i\pi^2}{3} + O (\epsilon) \, , \nonumber \\
P.P. \left(\int d^n\bar{q} \frac{q_{\mu}q_{\nu}q_{\rho}q_{\sigma}}
{\db{i}\db{j}\db{k}\db{l}}
\right)  & = &
- \frac{i\pi^2}{12 \epsilon}\,
\Bigl(g_{\mu\nu}g_{\rho\sigma} + g_{\mu\rho}g_{\nu\sigma} +
g_{\mu\sigma}g_{\nu\rho}\Bigr) \, .
\end{eqnarray}
\section{QCD Feyman Rules and diagrams \label{appb}}
In this appendix, we present
the Feynman rules and the diagrams used in the calculation.
In fig.~\ref{qcdrules} we list QCD propagators and vertices
as well as our parametrization of the $Vqq$ and $Sqq$ couplings.
Ghosts are drawn with dashed arrows, vectors with wavy lines and
scalars with dotted lines; Greek letters denote Lorentz indices;
$k,l= 1,2,3$ are the three colors of the quarks while
all remaining color indices range from 1 to 8;
$f^{abc}$ is the QCD $SU(N_{col})$ structure constant and
$t^a$ $(a = 1, \cdots, 8)$ are the color matrices
in the fundamental representation;
$m_q$ is the quark mass and $V_{\mu_1\mu_2\mu_3}(p_1,p_2,p_3)$ is given by
\bqa
\label{eq:vdef}
V_{\mu_1\mu_2\mu_3}(p_1,p_2,p_3) = g_{\mu_1\mu_2}(p_2 - p_1)_{\mu_3} +
g_{\mu_2\mu_3}(p_3 - p_2)_{\mu_1} + g_{\mu_3\mu_1}(p_1 - p_3)_{\mu_2}\,.
\eqa
Finally, in fig.~\ref{loopdiagrams} and ~\ref{loopmixdiagrams}, we draw
the pure QCD graphs and the mixed QCD diagrams which give
a non vanishing contribution
to ${\rm R_2}$. As explained in section~\ref{origin}, diagrams involving QCD FP ghosts
do not to contribute to ${\rm R_2}$ and are not included.
\begin{figure}[ht]
\bfig(300,400)
\SetScale{0.5}
%
%
\sof(-40,340)
\gluon{35,50}{135,50}{9}
\LongArrow(75,62)(95,62)
\Text(40,34)[bl]{$p$}
\Text(17,15)[]{$\mu$,$a$}
\Text(68,15)[]{$\nu$,$b$}
\Text(80,25)[l]{$\displaystyle =- i\, \frac{g_{\mu\nu}}{p^2}\, \delta_{ab}$ ,}
%
%
\sof(115,340)
\LongArrow(75,62)(98,62)
\Text(40,34)[bl]{$p$}
\flin{35,50}{135,50}
\Text(17,17)[]{$l$}
\Text(68,17)[]{$k$}
\Text(80,25)[l]{$\displaystyle = \frac{i\,\delta_{kl}}{\rlap/p - m_q}$ ,}
%
%
\sof(260,340)
\LongArrow(75,62)(98,62)
\Text(40,34)[bl]{$p$}
\DashArrowLine(35,50)(135,50){5}
\Text(17,17)[]{$a$}
\Text(68,17)[]{$b$}
\Text(80,25)[l]{$\displaystyle = \frac{i\,\delta_{ab}}{p^2} \,$ ,}
%
%
\sof(-33,260)
\gluon{130,5}{85,50}{5}\gluon{85,50}{130,95}{5}
\gluon{20,50}{85,50}{5}
\LongArrow(35,62)(66,62)
\LongArrow(105,10)(85,30)
\LongArrow(110,95)(90,75)
\Text(46,49)[]{$p_2$}
\Text(25,40)[]{$p_1$}
\Text(45,3)[]{$p_3$}
\Text(75,55)[]{$\mu_2$,$a_2$}
\Text(7,15)[]{$\mu_1$,$a_1$}
\Text(75,-6)[]{$\mu_3$,$a_3$}
\Text(90,25)[l]{$ \displaystyle  =  g \,f^{a_1a_2a_3}\,
V_{\mu_1\mu_2\mu_3}(p_1,p_2,p_3)\,$ ,}
%
%
\sof(-30,180)
\gluon{130,5}{85,50}{5}\gluon{85,50}{130,95}{5}
\gluon{40,5}{85,50}{5}\gluon{85,50}{40,95}{5}
\Text(68,0)[l]{\small $\rho$,$c$}
\Text(17,0)[r]{\small $\sigma$,$d$}
\Text(68,50)[l]{\small $\nu$,$b$}
\Text(17,50)[r]{\small $\mu$,$a$}
\Text(90,25)[l]{$\displaystyle = - ig^2\,[
f^{ebc}f^{eda}(g_{\nu \sigma} g_{\mu \rho} - g_{\mu \nu} g_{\rho \sigma}) $}
\Text(123,0)[l]{$ \displaystyle +
f^{ebd}f^{eac}(g_{\mu \nu} g_{\rho \sigma} - g_{\nu \rho} g_{\mu \sigma}) $}
\Text(123,-25)[l]{$ \displaystyle +
f^{eba}f^{ecd}(g_{\nu \rho} g_{\mu \sigma} - g_{\nu \sigma} g_{\mu \rho})] $ ,}
%
%
\sof(-30,80)
\gluon{130,5}{85,50}{5}\flin{85,50}{130,95} \flin{20,50}{85,50}
\Text(70,55)[l]{$k$}
\Text(7,15)[l]{$l$}
\Text(75,-6)[]{$\mu$,$a$}
\Text(90,25)[l]{$\displaystyle = -i g t_{kl}^a \gamma^{\mu}$ ,}
%
%
\sof(200,80)
\gluon{130,5}{85,50}{5} \ghlin{85,50}{130,95} \ghlin{20,50}{85,50}
\Text(46,49)[]{$p$}
\LongArrow(110,95)(90,75)
\Text(70,55)[l]{$a$}
\Text(7,15)[l]{$b$}
\Text(75,-6)[]{$\mu$,$c$}
\Text(90,25)[l]{$\displaystyle =  g f^{abc} p_{\mu}$ .}
%
%
\sof(-30,0)
\glin{130,5}{85,50}{8}\flin{85,50}{130,95} \flin{20,50}{85,50}
\Text(70,55)[l]{$k$}
\Text(7,15)[l]{$l$}
\Text(75,-6)[]{$\mu$}
\Text(50, 5)[]{\small $V$}
\Text(90,25)[l]{$\displaystyle = \,\delta_{kl}\, \gamma_{\mu} (v +a \gamma_5)$}
%
%
\sof(200,0)
\DashLine(130,5)(85,50){2}
\flin{85,50}{130,95} \flin{20,50}{85,50}
\Text(70,55)[l]{$k$}
\Text(7,15)[l]{$l$}
\Text(50, 5)[]{\small $S$}
\Text(90,25)[l]{$\displaystyle = \,\delta_{kl}\,(c+d \gamma_5)$}
\end{picture}
\end{center}
\caption{\em Feynman rules used for the computation.
The last two vertices parametrize a generic coupling
of a (pseudo)-vector $V$ and of a (pseudo)-scalar $S$ with a quark line, respectively.}
\label{qcdrules}
\end{figure}
\begin{figure}[ht]
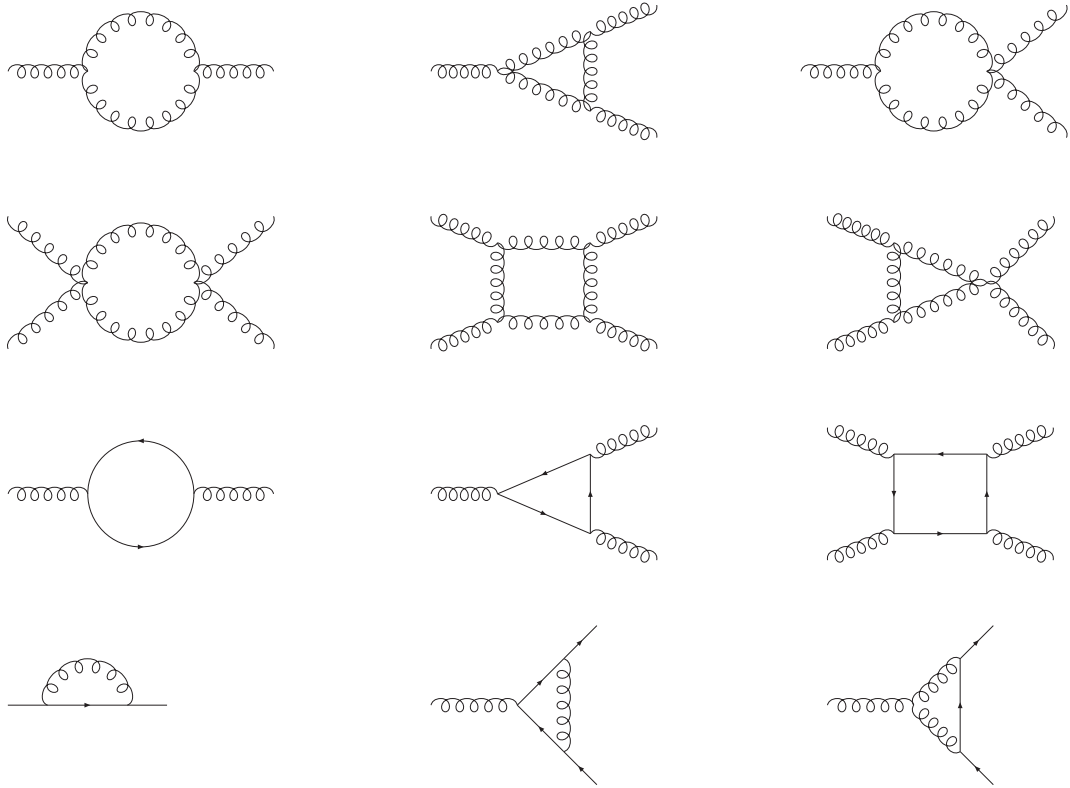

\bfig(300,320)
\SetScale{0.5}
\sof(-80,240)
\GlueArc(150,50)(40,0,180){5}{8}
\GlueArc(150,50)(40,180,360){5}{8}
\Gluon(50,50)(110,50){5}{5}
\Gluon(190,50)(250,50){5}{5}
\sof(70,240)
\Gluon(70,50)(120,50){5}{5}
\Gluon(190,80)(240,100){-5}{5}
\Gluon(190,20)(240,-0){5}{5}
\Gluon(190,20)(190,80){5}{5}
\Gluon(120,50)(190,80){-5}{6}
\Gluon(120,50)(190,20){5}{6}
\sof(220,240)
\GlueArc(150,50)(40,0,180){5}{8}
\GlueArc(150,50)(40,180,360){5}{8}
\Gluon(50,50)(110,50){5}{5}
\Gluon(190,50)(250,100){-5}{5}
\Gluon(190,50)(250,-0){5}{5}
\sof(-80,160)
\GlueArc(150,50)(40,0,180){5}{8}
\GlueArc(150,50)(40,180,360){5}{8}
\Gluon(110,50)(50,100){5}{5}
\Gluon(50,-0)(110,50){5}{5}
\Gluon(190,50)(250,100){-5}{5}
\Gluon(190,50)(250,-0){5}{5}
\sof(70,160)
\Gluon(120,80)(70,100){5}{5}
\Gluon(70,-0)(120,20){5}{5}
\Gluon(190,80)(240,100){-5}{5}
\Gluon(190,20)(240,-0){5}{5}
\Gluon(120,20)(120,80){-5}{5}
\Gluon(190,20)(190,80){5}{5}
\Gluon(120,80)(190,80){-5}{5}
\Gluon(120,20)(190,20){5}{5}
\sof(220,160)
\Gluon(120,80)(70,100){5}{6}
\Gluon(70,-0)(120,20){5}{5}
\Gluon(190,50)(240,100){-5}{6}
\Gluon(190,50)(240,-0){5}{6}
\Gluon(120,20)(120,80){-5}{5}
\Gluon(120,80)(190,50){-5}{6}
\Gluon(120,20)(190,50){5}{6}
\sof(-80,80)
\ArrowArc(150,50)(40,0,180)
\ArrowArc(150,50)(40,180,360)
\Gluon(50,50)(110,50){5}{5}
\Gluon(190,50)(250,50){5}{5}
\sof(70,80)
\Gluon(70,50)(120,50){5}{5}
\Gluon(190,80)(240,100){-5}{5}
\Gluon(190,20)(240,-0){5}{5}
\flin{190,80}{120,50}
\flin{120,50}{190,20}
\flin{190,20}{190,80}
\sof(220,80)
\Gluon(120,80)(70,100){5}{5}
\Gluon(70,-0)(120,20){5}{5}
\Gluon(190,80)(240,100){-5}{5}
\Gluon(190,20)(240,-0){5}{5}
\flin{120,80}{120,20}
\flin{190,20}{190,80}
\flin{190,80}{120,80}
\flin{120,20}{190,20}
\sof(-65,0)
\flin{20,50}{140,50}
\GlueArc(80,50)(30,0,180){5}{6}
\sof(95,0)
\flin{145,-10}{120,15}
\flin{120,15}{85,50}
\flin{85,50}{120,85}
\flin{120,85}{145,110}
\gluon{20,50}{85,50}{5}
\gluon{120,85}{120,15}{5}
\sof(245,0)
\flin{145,-10}{120,15}
\gluon{120,15}{85,50}{4}
\gluon{85,50}{120,85}{4}
\flin{120,85}{145,110}
\gluon{20,50}{85,50}{5}
\flin{120,15}{120,85}
\end{picture}
\end{center}
\caption{\em Pure QCD diagrams used for the calculation.}
\label{loopdiagrams}
\end{figure}
\begin{figure}[ht]
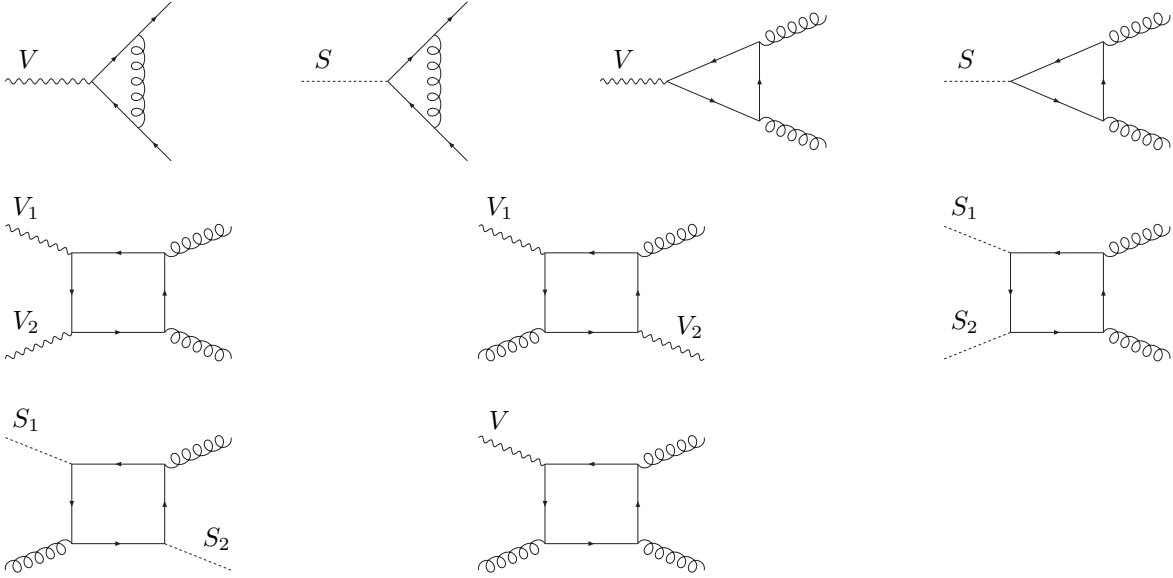

\bfig(300,240)
\SetScale{0.5}
\sof(-80,160)
\flin{145,-10}{120,15}
\flin{120,15}{85,50}
\flin{85,50}{120,85}
\flin{120,85}{145,110}
\glin{20,50}{85,50}{8}
\gluon{120,85}{120,15}{5}
\Text(15,30)[bl]{\small $V$}
\sof(32,160)
\flin{145,-10}{120,15}
\flin{120,15}{85,50}
\flin{85,50}{120,85}
\flin{120,85}{145,110}
\DashLine(20,50)(85,50){2}
\gluon{120,85}{120,15}{5}
\Text(15,30)[bl]{\small $S$}
\sof(120,160)
\glin{70,50}{120,50}{7}
\Gluon(190,80)(240,100){-5}{5}
\Gluon(190,20)(240,-0){5}{5}
\flin{190,80}{120,50}
\flin{120,50}{190,20}
\flin{190,20}{190,80}
\Text(40,30)[bl]{\small $V$}
\sof(250,160)
\DashLine(70,50)(120,50){2}
\Gluon(190,80)(240,100){-5}{5}
\Gluon(190,20)(240,-0){5}{5}
\flin{190,80}{120,50}
\flin{120,50}{190,20}
\flin{190,20}{190,80}
\Text(40,30)[bl]{\small $S$}
\sof(-105,80)
\glin{120,80}{70,100}{7}
\glin{70,0}{120,20}{7}
\Gluon(190,80)(240,100){-5}{5}
\Gluon(190,20)(240,-0){5}{5}
\flin{120,80}{120,20}
\flin{190,20}{190,80}
\flin{190,80}{120,80}
\flin{120,20}{190,20}
\Text(43,57)[]{\small $V_1$}
\Text(43,14)[]{\small $V_2$}
\sof(74,80)
\glin{120,80}{70,100}{7}
\Gluon(70,-0)(120,20){5}{5}
\Gluon(190,80)(240,100){-5}{5}
\glin{190,20}{240,0}{7}
\flin{120,80}{120,20}
\flin{190,20}{190,80}
\flin{190,80}{120,80}
\flin{120,20}{190,20}
\Text(43,57)[]{\small $V_1$}
\Text(115,12)[]{\small $V_2$}
\sof(250,80)
\DashLine(120,80)(70,100){2}
\DashLine(70,0)(120,20){2}
\Gluon(190,80)(240,100){-5}{5}
\Gluon(190,20)(240,-0){5}{5}
\flin{120,80}{120,20}
\flin{190,20}{190,80}
\flin{190,80}{120,80}
\flin{120,20}{190,20}
\Text(43,57)[]{\small $S_1$}
\Text(43,14)[]{\small $S_2$}
\sof(-105,0)
\DashLine(120,80)(70,100){2}
\Gluon(70,-0)(120,20){5}{5}
\Gluon(190,80)(240,100){-5}{5}
\DashLine(190,20)(240,0){2}
\flin{120,80}{120,20}
\flin{190,20}{190,80}
\flin{190,80}{120,80}
\flin{120,20}{190,20}
\Text(43,57)[]{\small $S_1$}
\Text(115,12)[]{\small $S_2$}
\sof(74,0)
\glin{120,80}{70,100}{7}
\Gluon(70,-0)(120,20){5}{5}
\Gluon(190,80)(240,100){-5}{5}
\Gluon(190,20)(240,-0){5}{5}
\flin{120,80}{120,20}
\flin{190,20}{190,80}
\flin{190,80}{120,80}
\flin{120,20}{190,20}
\Text(43,57)[]{\small $V$}
\end{picture}
\end{center}
\caption{\em Mixed QCD diagrams contributing to ${\rm R_2}$.}
\label{loopmixdiagrams}
\end{figure}

\section{Effective vertices in the color connection language \label{appc}}
In this appendix, we write down the Feynman rules for the QCD effective vertices
in the color connection language. Such rules are obtained by contracting
any gluon index $a_i$, appearing in the vertices
of figs.~\ref{effectivevertices} and~\ref{effectivemixvertices}
, by a color matrix $t^{a_i}_{k_i l_i}$.
Any gluonic color index  $a_i$ is therefore
projected out in terms of two quark like color and anti-color indices
${k_i}$ and ${l_i}$.
By then summing over gluon indices with the rule
\bqa
t^{a}_{kl} t^{a}_{ij} = \frac{1}{2}
\left[ \delta_{kj} \delta_{il} - \frac{1}{N_{col}}\delta_{kl}\delta_{ij}
\right]\,,
\eqa
the color part of the effective vertices can be entirely written down
in terms of $\delta$'s, which correspond to color connections.
Graphically, a color connection can be represented with a solid line, in
such a way that two solid lines stand for a gluon, while one single solid
line symbolize a quark. Finally, different color lines
can be connected by the exchange  of a scalar colorless gluon,
represented by a dashed line.
In such a language, the pure QCD effective
vertices of fig.~\ref{effectivevertices}
can be written as in figs.~\ref{colorconnections4}
and~\ref{colorconnections32}. Analogously, the last five mixed QCD vertices
of fig.~\ref{effectivemixvertices} give the results reported
in figs.~\ref{colorconnectionsmix4}-\ref{colorconnectionsmix3}.
\begin{center}
\begin{figure}[ht]
\bfig(300,455)
\SetScale{0.5}
\sof(-60,397)
\lin{130,105}{85,60}
\lin{140,95}{95,50}
\lin{140,5}{95,50}
\lin{130,-5}{85,40}
\lin{40,105}{85,60}
\lin{30,95}{75,50}
\lin{30,5}{75,50}
\lin{40,-5}{85,40}
\GCirc(85,50){4}{0}
\Text(10,55)[]{$i$}
\Text(75,55)[]{$j$}
\Text(75,-5)[]{$k$}
\Text(10,-5)[]{$l$}
\Text(70,25)[l]{
$
\displaystyle = \frac{i g^4 N_{col}}{192\pi^2}\,
\left[ \left(\frac{5}{4}+\frac{\lambda_{HV}}{2}
+\frac{3}{2} \frac{N_f}{N_{col}} \right) (g_{\mu_i\mu_j} g_{\mu_k\mu_l}
+g_{\mu_i\mu_l} g_{\mu_j\mu_k}) \right.$
}
\Text(130,-10)[l]{
$
\displaystyle \left.
-\,\left(3+\lambda_{HV}
+\frac{5}{2} \frac{N_f}{N_{col}} \right )\,g_{\mu_i\mu_k} g_{\mu_j\mu_l}
 \right]$
}
\sof(-60,304)
\lin{130,95}{85,50}
\lin{130,5}{85,50}
\lin{40,95}{85,50}
\lin{40,5.}{85,50}
\lin{140,85}{105,50}
\lin{140,15}{105,50}
\lin{30,85}{65,50}
\lin{30,15.}{65,50}
\GCirc(85,50){4}{0}
\Text(10,50)[]{$i$}
\Text(75,50)[]{$j$}
 \Text(75,0)[]{$k$}
 \Text(10,0)[]{$l$}
\Text(70,25)[l]{
$
\displaystyle = -\frac{i g^4}{192\pi^2}\,
(
 g_{\mu_i\mu_j} g_{\mu_3\mu_l}
+g_{\mu_i\mu_l} g_{\mu_j\mu_k}
+g_{\mu_i\mu_k} g_{\mu_j\mu_l}
)$}
\sof(-60,211)
\lin{130,95}{85,50}
\lin{130,5}{85,50}
\zlin{60,75}{85,50}
\zlin{60,25}{85,50}
\lin{40,25}{60,25}
\lin{60,5}{60,25}
\lin{140,85}{105,50}
\lin{140,15}{105,50}
\lin{40,75}{60,75}
\lin{60,95}{60,75}
\lin{40,75}{30,85} %
\lin{60,95}{50,105}
\lin{50,-5}{60,5}
\lin{40,25}{30,15}
\GCirc(85,50){4}{0}
\Text(70,25)[l]{$\displaystyle= - \frac{N_f}{N^2_{col}}$}
\sof(45,211)
\lin{130,95}{85,50}
\lin{130,5}{85,50}
\lin{40,95}{85,50}
\lin{40,5.}{85,50}
\lin{140,85}{105,50}
\lin{140,15}{105,50}
\lin{30,85}{65,50}
\lin{30,15.}{65,50}
\GCirc(85,50){4}{0}
\sof(-60,118)
\lin{140,15}{130,25}
\lin{110,5}{120,-5}
\lin{110,25}{130,25}
\lin{110,25}{110,5}
\lin{130,75}{140,85}
\lin{110,95}{120,105}
\lin{110,75}{130,75}
\lin{110,75}{110,95}
\zlin{110,75}{85,50}
\zlin{110,25}{85,50}
\zlin{60,75}{85,50}
\zlin{60,25}{85,50}
\lin{40,25}{60,25}
\lin{60,5}{60,25}
\lin{40,75}{60,75}
\lin{60,95}{60,75}
\lin{50,-5}{60,5}
\lin{40,25}{30,15}
\lin{40,75}{30,85}
\lin{60,95}{50,105}
\GCirc(85,50){4}{0}
\Text(70,25)[l]{$\displaystyle= 3 \frac{N_f}{N^3_{col}}$}
\sof(45,118)
\lin{130,95}{85,50}
\lin{130,5}{85,50}
\lin{40,95}{85,50}
\lin{40,5.}{85,50}
\lin{140,85}{105,50}
\lin{140,15}{105,50}
\lin{30,85}{65,50}
\lin{30,15.}{65,50}
\GCirc(85,50){4}{0}
\sof(-60,25)
\lin{130,105}{85,60}
\lin{140,95}{95,50}
\lin{140,5}{95,50}
\lin{130,-5}{85,40}
\lin{40,105}{85,60}
\lin{30,95}{95,30}
\zlin{60,25}{85,50}
\lin{40,25}{60,25}
\lin{60,5}{60,25}
\lin{50,-5}{60,5}
\lin{40,25}{30,15}
\GCirc(85,50){4}{0}
\Text(70,25)[l]{$\displaystyle= - \frac{1}{2}\,
\left(1-\frac{N_f}{N_{col}}
\right)
$}
\sof(95,25)
\lin{130,95}{85,50}
\lin{130,5}{85,50}
\lin{40,95}{85,50}
\lin{40,5.}{85,50}
\lin{140,85}{105,50}
\lin{140,15}{105,50}
\lin{30,85}{65,50}
\lin{30,15.}{65,50}
\GCirc(85,50){4}{0}
\end{picture}
\end{center}
\caption{\em Effective 4-gluon vertices contributing to ${\rm R_2}$ in pure QCD
in the color connection language. $N_f$ is the number of fermions running
in the quark loop.}
\label{colorconnections4}
\end{figure}
\end{center}
\begin{center}
\begin{figure}[ht]
\bfig(300,430)
\SetScale{0.5}
\sof(-60,380)
\lin{130,105}{82.5,57.5}
\lin{140,95}{95,50}
\lin{140,5}{95,50}
\lin{130,-5}{82.5,42.5}
\lin{22,57.5}{82.5,57.5}
\lin{22,42.5}{82.5,42.5}
\GCirc(85,50){4}{0}
\Text(3,24)[]{$i$}
\Text(78,55)[]{$j$}
\Text(78,-5)[]{$k$}
\Text(80,25)[l]{
$
\displaystyle = \frac{i g^3 N_{col}}{192\pi^2}\,
\left[ \left( \frac{7}{4}+\lambda_{HV}
\right) + 2 \frac{N_f}{N_{col}}\right]\,V_{\mu_i\mu_j\mu_k}(p_i,p_j,p_j)
$}
\sof(-60,310)
\lin{22,57.5}{142.5,57.5}
\lin{22,42.5}{142.5,42.5}
\Text(2,25)[]{$i$}
\Text(79,25)[]{$j$}
\GCirc(85,50){4}{0}
\Text(84,25)[l]{
$
\displaystyle = \frac{i g^2 N_{col}}{96\pi^2}\,
\, \left[\,\frac{p^2}{2} g_{\mu_i\mu_j}
+\lambda_{HV}\,\Bigl( g_{\mu_i\mu_j} p^2-p_{\mu_i} p_{\mu_j}\Bigr)\right.
$}
\Text(140,-10)[l]{$\displaystyle \left.
+\frac{N_f}{N_{col}}\,(p^2-6\, m_q^2)\,g_{\mu_i\mu_j}\right]$}
\sof(-60,220)
\lin{22,58}{37,58}
\lin{22,42}{37,42}
\zlin{45,50}{120,50}
\lin{128,58}{143,58}
\lin{128,42}{143,42}
\lin{127,42}{120,50}
\lin{120,50}{128,58}
\lin{37,58}{45,50}
\lin{37,42}{45,50}
\GCirc(85,50){4}{0}
\Text(80,25)[l]{$\displaystyle = -\frac{1}{N_{col}}$}
\sof(62,220)
\lin{22,57.5}{142.5,57.5}
\lin{22,42.5}{142.5,42.5}
\GCirc(85,50){4}{0}
\sof(-60,150)
\lin{130,105}{82.5,57.5}
\lin{130,-5}{82.5,42.5}
\lin{22,57.5}{82.5,57.5}
\lin{22,42.5}{82.5,42.5}
\Text(2,25)[]{$\mu$}
\GCirc(85,50){4}{0}
\Text(80,25)[l]{$\displaystyle = \frac{i g^3}{32 \pi^2 }
\frac{N_{col}^2-1}{2 N_{col}}\,\gamma_\mu\,(1+ \lambda_{HV})$}
\sof(-60,80)
\lin{22,58}{37,58}
\lin{22,42}{37,42}
\zlin{45,50}{90,50}
\lin{37,58}{45,50}
\lin{37,42}{45,50}
\lin{140,95}{95,50}
\lin{140,5}{95,50}
\GCirc(90,50){4}{0}
\Text(80,25)[l]{$\displaystyle = -\frac{1}{N_{col}}$}
\sof(62,80)
\lin{130,105}{82.5,57.5}
\lin{130,-5}{82.5,42.5}
\lin{22,57.5}{82.5,57.5}
\lin{22,42.5}{82.5,42.5}
\GCirc(85,50){4}{0}
\sof(-60,10)
\lin{22,50}{142.5,50}
\GCirc(85,50){4}{0}
\Text(80,25)[l]{$\displaystyle = \frac{i g^2}{16 \pi^2}
\frac{N_{col}^2-1}{2 N_{col}}
(-\rlap/p + 2 m_q)\,\lambda_{HV}$}
\end{picture}
\end{center}
\caption{\em Effective 3- and 2- point vertices
contributing to ${\rm R_2}$ in pure QCD
in the color connection language. All momenta are incoming.
The first three diagrams represent the
$ggg$ and $gg$ vertices; the last three $qqg$ and $qq$.
$N_f$ is the number of fermions running
in the quark loop.}
\label{colorconnections32}
\end{figure}
\end{center}
\begin{center}
\begin{figure}[ht]
\bfig(300,280)
\SetScale{0.5}
\sof(-60,210)
\lin{130,105}{85,60}
\lin{140,95}{95,50}
\lin{140,5}{95,50}
\lin{130,-5}{85,40}
\lin{40,105}{85,60}
\lin{30,95}{95,30}
\glin{35,0}{85,50}{7}
\Text(12,-3)[r]{$\mu$}
\Text(10,55)[]{$i$}
\Text(75,55)[]{$j$}
\Text(75,-5)[]{$k$}
\GCirc(85,50){4}{0}
\Text(80,25)[l]{$\displaystyle = -\frac{g^3}{64\pi^2}
\left [\,v\,\frac{1}{3}
(g_{\mu\alpha_i} g_{\alpha_j\alpha_k}
+g_{\mu\alpha_j} g_{\alpha_i\alpha_k}
+g_{\mu\alpha_k} g_{\alpha_i\alpha_j})
\,-i3a\,\epsilon_{\mu\alpha_i\alpha_j\alpha_k}
\right]$}
\sof(-60,120)
\lin{130,95}{85,50}
\lin{130,5}{85,50}
\zlin{60,75}{85,50}
\glin{35,0}{85,50}{7}
\lin{140,85}{105,50}
\lin{140,15}{105,50}
\lin{40,75}{60,75}
\lin{60,95}{60,75}
\lin{40,75}{30,85} %
\lin{60,95}{50,105}
\Text(12,-3)[r]{$\mu$}
\Text(10,50)[]{$i$}
\Text(75,50)[]{$j$}
\Text(75,0)[]{$k$}
\GCirc(85,50){4}{0}
\Text(80,25)[l]{$\displaystyle = \frac{g^3}{96\pi^2 N_{col}}\,v\,
(g_{\mu\alpha_i} g_{\alpha_j\alpha_k}
+g_{\mu\alpha_j} g_{\alpha_i\alpha_k}
+g_{\mu\alpha_k} g_{\alpha_i\alpha_j})$}
\sof(-60,30)
\lin{140,15}{130,25}
\lin{110,5}{120,-5}
\lin{110,25}{130,25}
\lin{110,25}{110,5}
\lin{130,75}{140,85}
\lin{110,95}{120,105}
\lin{110,75}{130,75}
\lin{110,75}{110,95}
\zlin{110,75}{85,50}
\zlin{110,25}{85,50}
\zlin{60,75}{85,50}
\glin{35,0}{85,50}{7}
\lin{40,75}{60,75}
\lin{60,95}{60,75}
\lin{40,75}{30,85}
\lin{60,95}{50,105}
\Text(12,-3)[r]{$\mu$}
\GCirc(85,50){4}{0}
\Text(80,25)[l]{$\displaystyle = -\frac{2}{N_{col}}$}
\sof(55,30)
\lin{130,95}{85,50}
\lin{130,5}{85,50}
\zlin{60,75}{85,50}
\glin{35,0}{85,50}{7}
\lin{140,85}{105,50}
\lin{140,15}{105,50}
\lin{40,75}{60,75}
\lin{60,95}{60,75}
\lin{40,75}{30,85} %
\lin{60,95}{50,105}
\Text(12,-3)[r]{$\mu$}
\GCirc(85,50){4}{0}
\end{picture}
\end{center}
\caption{\em Effective Vggg vertex contributing to ${\rm R_2}$ in mixed QCD
in the color connection language (contribution of one quark loop).}
\label{colorconnectionsmix4}
\end{figure}
\end{center}
\begin{center}
\begin{figure}[ht]
\bfig(300,185)
\SetScale{0.5}
\sof(-78,110)
\lin{130,95}{85,50}
\lin{130,5}{85,50}
\glin{35,0}{85,50}{7}
\glin{35,100}{85,50}{7}
\lin{140,85}{105,50}
\lin{140,15}{105,50}
\Text(13,-3)[r]{$\mu_1$}
\Text(13,53)[r]{$\mu_2$}
\Text(75,50)[]{$i$}
\Text(75,0)[]{$j$}
\GCirc(85,50){4}{0}
\Text(70,25)[l]{$\displaystyle = -{N_{col}}$}
\sof(15,110)
\lin{140,15}{130,25}
\lin{110,5}{120,-5}
\lin{110,25}{130,25}
\lin{110,25}{110,5}
\lin{130,75}{140,85}
\lin{110,95}{120,105}
\lin{110,75}{130,75}
\lin{110,75}{110,95}
\zlin{110,75}{85,50}
\zlin{110,25}{85,50}
\glin{35,0}{85,50}{7}
\glin{35,100}{85,50}{7}
\Text(13,-3)[r]{$\mu_1$}
\Text(13,53)[r]{$\mu_2$}
\Text(73,50)[]{$i$}
\Text(73,0)[]{$j$}
\GCirc(85,50){4}{0}
\Text(65,25)[l]{$\displaystyle = -\frac{i g^2}{48 \pi^2}
\,(v_1 v_2+a_1 a_2)\,
( g_{\mu_1\mu_2} g_{\alpha_i\alpha_j}
 +g_{\mu_1\alpha_i} g_{\mu_2\alpha_j}
 +g_{\mu_1\alpha_j} g_{\mu_2\alpha_i})
$}
\sof(-78,10)
\lin{130,95}{85,50}
\lin{130,5}{85,50}
\DashLine(35,0)(85,50){2}
\DashLine(35,100)(85,50){2}
\lin{140,85}{105,50}
\lin{140,15}{105,50}
\Text(73,50)[]{$i$}
\Text(73,0)[]{$j$}
\GCirc(85,50){4}{0}
\Text(70,25)[l]{$\displaystyle =-{N_{col}}$}
\sof(15,10)
\lin{140,15}{130,25}
\lin{110,5}{120,-5}
\lin{110,25}{130,25}
\lin{110,25}{110,5}
\lin{130,75}{140,85}
\lin{110,95}{120,105}
\lin{110,75}{130,75}
\lin{110,75}{110,95}
\zlin{110,75}{85,50}
\zlin{110,25}{85,50}
\DashLine(35,0)(85,50){2}
\DashLine(35,100)(85,50){2}
\Text(73,50)[]{$i$}
\Text(73,0)[]{$j$}
\GCirc(85,50){4}{0}
\Text(75,25)[l]{$\displaystyle = \frac{i g^2}{16 \pi^2}
\,(c_1 c_2-d_1 d_2)\,g_{\alpha_i \alpha_j}$}
\end{picture}
\end{center}
\caption{\em Effective VVgg and SSgg vertices contributing
to ${\rm R_2}$ in mixed QCD
in the color connection language.
In the case of neutral external vectors or scalars, the formulae should 
be read as the contribution given by one quark loop. 
In the case of charged external particles, they refer instead 
to the contribution of one quark family.}
\label{colorconnectionsmix4b}
\end{figure}
\end{center}

\begin{center}
\begin{figure}[ht]
\bfig(300,180)
\SetScale{0.5}
\sof(-78,110)
\lin{130,95}{85,50}
\lin{130,5}{85,50}
\glin{25,50}{85,50}{7}
\lin{140,85}{105,50}
\lin{140,15}{105,50}
\Text(10,28)[r]{$\mu$}
\Text(73,50)[]{$i$}
\Text(73,0)[]{$j$}
\GCirc(85,50){4}{0}
\Text(70,25)[l]{$\displaystyle = -{N_{col}}$}
\sof(37,110)
\lin{140,15}{130,25}
\lin{110,5}{120,-5}
\lin{110,25}{130,25}
\lin{110,25}{110,5}
\lin{130,75}{140,85}
\lin{110,95}{120,105}
\lin{110,75}{130,75}
\lin{110,75}{110,95}
\zlin{110,75}{85,50}
\zlin{110,25}{85,50}
\glin{25,50}{85,50}{7}
\Text(10,28)[r]{$\mu$}
\Text(73,50)[]{$i$}
\Text(73,0)[]{$j$}
\GCirc(85,50){4}{0}
\Text(65,25)[l]{
$\displaystyle =- a \,\frac{ig^2}{24 \pi^2}
\epsilon_{\mu\alpha_i\alpha_j\beta}\, (p_i-p_j)^{\beta}$}
\sof(-78,10)
\lin{130,95}{85,50}
\lin{130,5}{85,50}
\DashLine(25,50)(85,50){2}
\lin{140,85}{105,50}
\lin{140,15}{105,50}
\Text(73,50)[]{$i$}
\Text(73,0)[]{$j$}
\GCirc(85,50){4}{0}
\Text(70,25)[l]{$\displaystyle =-{N_{col}}$}
\sof(37,10)
\lin{140,15}{130,25}
\lin{110,5}{120,-5}
\lin{110,25}{130,25}
\lin{110,25}{110,5}
\lin{130,75}{140,85}
\lin{110,95}{120,105}
\lin{110,75}{130,75}
\lin{110,75}{110,95}
\zlin{110,75}{85,50}
\zlin{110,25}{85,50}
\DashLine(25,50)(85,50){2}
\Text(73,50)[]{$i$}
\Text(73,0)[]{$j$}
\GCirc(85,50){4}{0}
\Text(75,25)[l]{$\displaystyle = c \,\frac{g^2}{8 \pi^2}
\,g_{\alpha_i \alpha_j}\,m_q$}
\end{picture}
\end{center}
\caption{\em Effective Vgg and Sgg vertices contributing
to ${\rm R_2}$ in mixed QCD
in the color connection language (contribution of one quark loop).
All momenta are incoming.}
\label{colorconnectionsmix3}
\end{figure}
\end{center}


\begin{thebibliography}{999}
%
\bibitem{Passarino:1978jh}
  G.~Passarino and M.~J.~G.~Veltman,
  Nucl.\ Phys.\  B {\bf 160} (1979) 151.
%
\bibitem{Denner:1991kt}
  A.~Denner,
  Fortsch.\ Phys.\  {\bf 41} (1993) 307
  [arXiv:0709.1075 [hep-ph]].
%
\bibitem{classic_techniques}
  J.~Kublbeck, M.~Bohm and A.~Denner,
  Comput.\ Phys.\ Commun.\  {\bf 60} (1990) 165;
\\
  R.~Mertig, M.~Bohm and A.~Denner,
  Comput.\ Phys.\ Commun.\  {\bf 64} (1991) 345;
\\
  A.~Denner and S.~Dittmaier,
  Nucl.\ Phys.\  B {\bf 658} (2003) 175
  [arXiv:hep-ph/0212259];
\\
  A.~Ferroglia, M.~Passera, G.~Passarino and S.~Uccirati,
  Nucl.\ Phys.\  B {\bf 650} (2003) 162
  [arXiv:hep-ph/0209219]; \\
  T.~Diakonidis, J.~Fleischer, J.~Gluza, K.~Kajda, T.~Riemann and J.~B.~Tausk,
  arXiv:0812.2134 [hep-ph]. 
%
\bibitem{golem}
   T.~Binoth, J.~P.~Guillet, G.~Heinrich, E.~Pilon and C.~Schubert,
   JHEP {\bf 0510}, 015 (2005) [arXiv:hep-ph/0504267]; \\
   T.~Binoth, J.~P.~Guillet, G.~Heinrich, E.~Pilon and T.~Reiter,
   arXiv:0810.0992 [hep-ph]; \\
   T.~Binoth {\it et al.},
   arXiv:0807.0605 [hep-ph].
%
\bibitem{classic_results}
  A.~Denner, S.~Dittmaier, M.~Roth and D.~Wackeroth,
  Phys.\ Lett.\  B {\bf 475} (2000) 127
  [arXiv:hep-ph/9912261];
\\
  G.~Montagna, F.~Piccinini, O.~Nicrosini, G.~Passarino and R.~Pittau,
  Comput.\ Phys.\ Commun.\  {\bf 76} (1993) 328;
\\
  D.~Y.~Bardin, P.~Christova, M.~Jack, L.~Kalinovskaya, A.~Olchevski, S.~Riemann and T.~Riemann,
  Comput.\ Phys.\ Commun.\  {\bf 133} (2001) 229
  [arXiv:hep-ph/9908433];
\\
  T.~Hahn and M.~Perez-Victoria,
  Comput.\ Phys.\ Commun.\  {\bf 118} (1999) 153
  [arXiv:hep-ph/9807565].

\bibitem{unitarity-cut}
  Z.~Bern, L.~J.~Dixon, D.~C.~Dunbar and D.~A.~Kosower,
  Nucl.\ Phys.\ B {\bf 435} (1995) 59
  [arXiv:hep-ph/9409265] and
  Nucl.\ Phys.\ B {\bf 425}, 217 (1994). See also Z.~Bern, 
  L.~J.~Dixon and D.~A.~Kosower,
  Annals Phys.\  {\bf 322} (2007) 1587
  [arXiv:0704.2798 [hep-ph]].

\bibitem{unitarity-results}
  Z.~Bern, L.~J.~Dixon and D.~A.~Kosower,
  Phys.\ Rev.\ Lett.\  {\bf 70} (1993) 2677
  [arXiv:hep-ph/9302280];
\\
  Z.~Bern, L.~J.~Dixon and D.~A.~Kosower,
  Nucl.\ Phys.\  B {\bf 513} (1998) 3
  [arXiv:hep-ph/9708239];
\\
  Z.~Bern, L.~J.~Dixon and D.~A.~Kosower,
  Nucl.\ Phys.\  B {\bf 437} (1995) 259
  [arXiv:hep-ph/9409393];
\\
  C.~F.~Berger, Z.~Bern, L.~J.~Dixon, D.~Forde and D.~A.~Kosower,
  Phys.\ Rev.\  D {\bf 74} (2006) 036009
  [arXiv:hep-ph/0604195].
%
\bibitem{Britto:2004nc}
  R.~Britto, F.~Cachazo and B.~Feng,
  Nucl.\ Phys.\ B {\bf 725}, 275 (2005).
%
\bibitem{opp}
  G.~Ossola, C.~G.~Papadopoulos and R.~Pittau,
  Nucl.\ Phys.\  B {\bf 763} (2007) 147 [arXiv:hep-ph/0609007].
%
\bibitem{Ossola:2007ax}
G.~Ossola, C.~G.~Papadopoulos and R.~Pittau,
  JHEP {\bf 0803} (2008) 042
  [arXiv:0711.3596 [hep-ph]]. 
  See also P.~Mastrolia, G.~Ossola, C.~G.~Papadopoulos and R.~Pittau,
  JHEP {\bf 0806} (2008) 030
  [arXiv:0803.3964 [hep-ph]].
%
\bibitem{intlevel}
  F.~del Aguila and R.~Pittau,
  JHEP {\bf 0407} (2004) 017
  [arXiv:hep-ph/0404120] and
  R.~Pittau,
  arXiv:hep-ph/0406105.
%
\bibitem{Ellis:2007br}
  R.~K.~Ellis, W.~T.~Giele and Z.~Kunszt,
  JHEP {\bf 0803} (2008) 003
  [arXiv:0708.2398 [hep-ph]].
%
\bibitem{Berger:2008sj}
  C.~F.~Berger {\it et al.},
  Phys.\ Rev.\  D {\bf 78}, 036003 (2008)
  [arXiv:0803.4180 [hep-ph]]; \\
  W.~B.~Kilgore,
  arXiv:0711.5015 [hep-ph].
%
\bibitem{Giele:2008ve}
  W.~T.~Giele, Z.~Kunszt and K.~Melnikov,
  JHEP {\bf 0804} (2008) 049
  [arXiv:0801.2237 [hep-ph]].
%
\bibitem{Lazopoulos:2008ex}
  A.~Lazopoulos,
  arXiv:0812.2998 [hep-ph].
%
\bibitem{directcomp1}
  T.~Binoth, J.~P.~Guillet and G.~Heinrich,
  JHEP {\bf 0702} (2007) 013
  [arXiv:hep-ph/0609054]; \\
  A.~Bredenstein, A.~Denner, S.~Dittmaier and S.~Pozzorini,
the  
  JHEP {\bf 0808} (2008) 108
  [arXiv:0807.1248 [hep-ph]].
%
\bibitem{directcomp2}
  Z.~G.~Xiao, G.~Yang and C.~J.~Zhu,
 Nucl.\ Phys.\  B {\bf 758} (2006) 1
  [arXiv:hep-ph/0607015]; \\
X.~Su, Z.~G.~Xiao, G.~Yang and C.~J.~Zhu,
  Nucl.\ Phys.\  B {\bf 758} (2006) 35
  [arXiv:hep-ph/0607016].
%
\bibitem{dcut}
  Z.~Bern and A.~G.~Morgan,
  Nucl.\ Phys.\  B {\bf 467}, 479 (1996)
  [arXiv:hep-ph/9511336]; \\
   Z.~Bern, L.~J.~Dixon, D.~C.~Dunbar and D.~A.~Kosower,
  Phys.\ Lett.\  B {\bf 394}, 105 (1997)
  [arXiv:hep-th/9611127]; \\
  C.~Anastasiou, R.~Britto, B.~Feng, Z.~Kunszt and P.~Mastrolia,
  Phys.\ Lett.\  B {\bf 645}, 213 (2007)
  [arXiv:hep-ph/0609191]; \\
  R.~Britto and B.~Feng,
  JHEP {\bf 0802}, 095 (2008)
  [arXiv:0711.4284 [hep-ph]]; \\
  R.~Britto, B.~Feng and P.~Mastrolia,
  Phys.\ Rev.\  D {\bf 78}, 025031 (2008)
  [arXiv:0803.1989 [hep-ph]]; \\
  R.~K.~Ellis, W.~T.~Giele, Z.~Kunszt and K.~Melnikov,
  arXiv:0806.3467 [hep-ph].
%
\bibitem{rec}
  Z.~Bern, L.~J.~Dixon and D.~A.~Kosower,
  Phys.\ Rev.\  D {\bf 71}, 105013 (2005)
  [arXiv:hep-th/0501240],
  Phys.\ Rev.\  D {\bf 72}, 125003 (2005)
  [arXiv:hep-ph/0505055] and
  Phys.\ Rev.\  D {\bf 73}, 065013 (2006)
  [arXiv:hep-ph/0507005];\\
  C.~F.~Berger, Z.~Bern, L.~J.~Dixon, D.~Forde and D.~A.~Kosower,
  Phys.\ Rev.\  D {\bf 74}, 036009 (2006)
  [arXiv:hep-ph/0604195].
%
\bibitem{Ossola:2008xq}
  G.~Ossola, C.~G.~Papadopoulos and R.~Pittau,
  JHEP {\bf 0805} (2008) 004
  [arXiv:0802.1876 [hep-ph]] and
  JHEP {\bf 0707} (2007) 085
  [arXiv:0704.1271 [hep-ph]].
%
\bibitem{vanHameren:2009dr}
  A.~van Hameren, C.~G.~Papadopoulos and R.~Pittau,
  arXiv:0903.4665 [hep-ph].
%
\bibitem{Kanaki:2000ey}
  A.~Kanaki and C.~G.~Papadopoulos,
  Comput.\ Phys.\ Commun.\  {\bf 132} (2000) 306
  [arXiv:hep-ph/0002082]; \\
  A.~Kanaki and C.~G.~Papadopoulos,
  arXiv:hep-ph/0012004.
\\
  A.~Cafarella, C.~G.~Papadopoulos and M.~Worek,
  arXiv:0710.2427 [hep-ph].
%
\bibitem{Giele:2008bc}
  W.~T.~Giele and G.~Zanderighi,
  arXiv:0805.2152 [hep-ph].
%
\bibitem{Bern:2008ef}
  Z.~Bern {\it et al.}  [NLO Multileg Working Group],
  arXiv:0803.0494 [hep-ph].
\end{thebibliography}
\end{document}